\documentclass[aps,prl,twocolumn,showpacs]{revtex4}
\usepackage{graphicx,dcolumn,bm,amsfonts,amsmath,amssymb}
\usepackage{mathptmx,longtable,float}

\begin{document}

\preprint{ver.1.0 / \today}
\title{Surface-termination dependent magnetism and strong perpendicular magnetocrystalline anisotropy
  of a FeRh (001) thin film: A density-functional study}
\author{Soyoung Jekal$^{1}$}
\author{S. H. Rhim$^{1}$}
\email[Email address: ]{sonny@ulsan.ac.kr}
\author{Soon Cheol Hong$^{1}$}
\email[Email address: ]{schong@ulsan.ac.kr}
\author{Won-joon Son$^{2}$}
\author{Alexander B. Shick$^{3}$}
\affiliation{
  $^{1}$Department of Physics and Energy Harvest-Storage Research Center, University of Ulsan,
  Ulsan, 680-749, Republic of Korea\\
  $^{2}$Samsung Advanced Institute of Technology, Suwon, 443-803, Republic of Korea\\
  $^{3}$Institute of Physics, ASCR, Na Slovance 2, CZ-18221 Prague, Czech Republic\\
}
\date{\today}

\begin{abstract}
  Magnetism of FeRh (001) films strongly depends on film thickness and surface terminations.
  While magnetic ground state of bulk FeRh is G-type antiferromagnetism,
  the Rh-terminated films exhibit ferromagnetism with strong perpendicular MCA
  whose energy +2.1 meV/$\Box$ is two orders of magnitude greater than bulk 3d magnetic metals,
  where $\Box$ is area of two-dimensional unit cell.
  While Goodenough-Kanamori-Anderson rule on the superexchange interaction is crucial
  in determining the magnetic ground phases of FeRh bulk and thin films,
  the magnetic phases are results of interplay and competition between three mechanisms
  – the superexchange interaction, the Zener-type direct-interaction, and magnetic energy gain. 
\end{abstract}

\pacs{68.37.Ef,75.70.Tj,75.70.Rf}
\maketitle
FeRh alloys have attracted significantly because of their various intriguing physics phenomena
including magneto-caloric effect and huge magnetoresistance\cite{1,2,3,4}.
Transition between antiferromagnetism (AFM) and ferromagnetism (FM) occurs
above room-temperature about 350 K and
ultrafast phase transition of magnetic phases in the FeRh alloys is induced
by femto-second laser, which have drawn more attention due to possibility of applications
for heat-assisted magnetic recording (HAMR)\cite{5,6,7}.

Furthermore, feasibility of room-temperature memories based on the AFM spintronics
has been successfully demonstrated in the FeRh alloys very recently
utilizing anisotropic magneto-resistance (AMR) of the bistable AFM states\cite{8,9,10}.
This AFM spintronics has some advantages over that based on FM states\cite{8,9,10,11,12,13,14,15,16}
owing to the absence of stray magnetic field from the zero net magnetization
and the insensitivity to the external magnetic fields.
Regarding FeRh thin films, several experiments have been reported\cite{17,18,19}.
Instead of the G-type antiferromagnetic (G-AFM) in bulk,
some thin films exhibit FM states stabilized at the interface with metal,
while the FM states are unstable at interface with oxide\cite{17}.
Interestingly, it was reported that an electric field of only a few volts is necessary
to drive the AFM-FM transition for the epitaxially grown FeRh films
on the ferroelectric BaTiO3 substrate\cite{18}.
Also, it was revealed that the spin orientation of the FeRh film on the MgO (001) depends
on the strength of lattice strain and magnetic state\cite{19}.

In this paper, magnetism and magnetocrystalline anisotropy (MCA) of FeRh films are investigated
using first-principles calculations. It is found that magnetism and MCA are significantly affected
not only by film thickness but also by the surface-terminations.
The Rh-terminated films are more stable in FM state by quite big energy differences
relative to the magnetic ground G-AFM state in bulk.
Furthermore, the Rh-termination exhibits strong perpendicular MCA of +2.1 meV/$\Box$,
where $\Box$ is two-dimensional (2D) unit cell area.
The Fe-terminations, on the other hand, are in G-AFM states as in bulk.
The strikingly different behavior of these two terminations is explained mainly
in the framework of Goodenough-Kanamori-Anderson (GKA) rule on the superexchange interaction\cite{20,21}
with the Zener-type direct exchange interaction\cite{22} also taken into account.

Density functional calculations are performed using Vienna \textit{ab initio} Simulation 
package (VASP)\cite{23}. Results by VASP, particularly MCA, have been 
double-checked with the highly precise full-potential linearized augmented plane 
wave (FLAPW) method\cite{24}. Generalized gradient approximation (GGA) within projector 
augmented-wave (PAW) scheme\cite{25} is employed for the exchange-correlation interaction. 
{\em k} meshes of 24$\times$24$\times$24 and 24$\times$24$\times$1 in Monkhorst-Pack scheme
are used for bulk and films, respectively. For wave function expansions, 500 eV is used for 
cutoff energy. Convergence with respect to cutoff energy and number of {\em k} 
points are seriously checked.
\begin{figure}
  \centering
  \includegraphics[width=8cm]{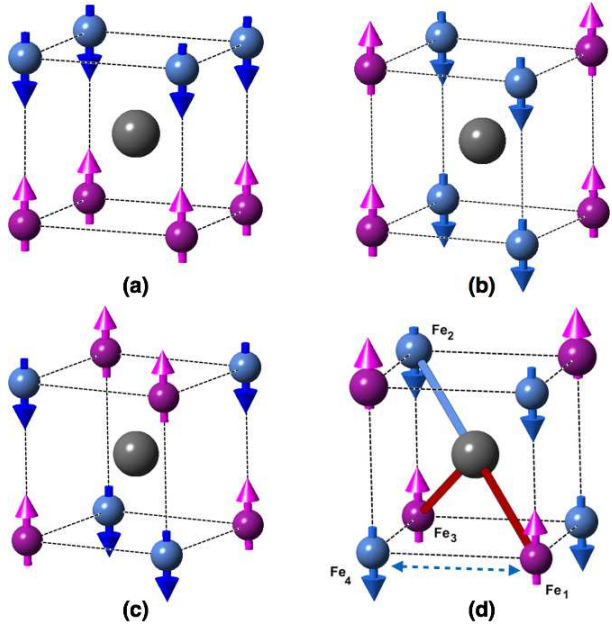}
  \caption{Schematic diagrams of magnetic structures of (a) A-, (b) C- and (c) G-AFM 
states of the bulk FeRh, and (d) exchange interaction between Fe atoms in the G-AFM 
state. Red and blue spheres at the corners represent Fe atoms with different magnetic states, respectively,
while grey spheres represent Rh atom.
In (d), cylinder connecting Fe$_1$ and Fe$_2$ denote the 180$^{\circ}$ superexchange,
while red cylinders connecting Fe$_1$-Rh-Fe$_3$ do the 90$^{\circ}$ superexchange.
Blue dotted line which connects Fe$_1$-Fe$_4$ denotes the direct exchange interaction.}
\label{fig:1}
\end{figure}
Magnetic structures of the bulk FeRh are illustrated schematically in Fig. 1(a)-(c) 
for A-, C- and G-type AFM, respectively, where arrows represent the direction of 
magnetic moments of Fe atoms. To account for the AFM structures, a tetragonal magnetic 
unit cell, $c(2\times2)$ in the {\em xy} plane and doubled along the {\em z} axis, 
is taken. To investigate thickness and surface-termination dependent magnetism 
of the FeRh (001) films, two terminations, Fe- and Rh-termination with thickness 
from 3- to 15-ML have been taken into account, where we allow relaxation along 
the {\em z}-axis while fixing two-dimensional (2D) lattice constant as the calculated 
bulk value (3.007 \AA) of the G-AFM.

%
%
%
First, magnetism of bulk FeRh is presented. AFM in G-type is found to be most stable 
from total energy calculations by 48.3, 56.7, and 187 meV/Fe with respect to FM 
and AFM in A- and C-type, respectively. Since the C-AFM has the highest energy, 
it will be excluded in forthcoming discussion if not necessary. Calculated lattice 
constant (3.012 \AA) of the bulk FeRh in FM state is a little bit larger than that 
in the G-AFM (3.007 \AA). Magnetic moments of Fe and Rh atoms in the G-AFM (FM) 
are 3.158 (3.144) $\mu_{B}$ and 0.00 (1.041) $\mu_{B}$, 
respectively. Lattice constants and magnetic moments in this work are reasonably 
consistent with experiments\cite{26,27} and previous calculation\cite{28}.

In FeRh (001) films, the interlayer spacing exhibits oscillatory feature: the topmost 
surface moves downward whereas the subsurface layers do upward, and so on, where 
the innermost layer converge to bulk behavior [See Supplementary Information for 
fully relaxed interlayer spacings of the Fe- and the Rh-terminated film in Fig. 
S1(a) and (b), respectively]. 7- or 9-ML films are thick enough due to short metallic 
screening length.

Total energy differences of the FM (A-AFM) and the G-AFM, which we denote as 
$\Delta E  = E_{FM} (E_{A}) - E_{G}$,
are presented in Fig. 2(a) and (b) for the Fe- and Rh-termination, respectively, 
where FM, A, and G in subscripts stand for FM, A-AFM, and G-AFM, respectively. 
As shown in Fig. 2(a), the G-AFM is the most stable for the Fe-termination regardless 
of thickness. We recall here that magnetism is so delicate physics phenomenon that 
it can behave differently in reduced dimension\cite{29,30}. As film gets thinner, $\Delta$E 
decreases while $\Delta$E=36.6 meV/Fe for 9-ML is still less than that of bulk 
(48.3 meV/Fe). As shown in Fig. 2(b), magnetism of the Rh-termination is totally 
different from the bulk. FM is favored over G-AFM up to 15-ML by big energy difference 
whose absolute value is larger than 70 meV/Fe for 7-ML.
Noteworthy, there is a crossover from FM to G-AFM when the number of layers exceeds 17-ML.
Furthermore, even A-AFM 
is more stable than G-AFM for films thinner than 7-ML. Magnetic moments of Fe and 
Rh atoms in both terminations are listed in Table 1 and 2 for their respective 
magnetic ground states, G-AFM and FM. The surface layers have almost the same magnetic moments
as the bulk unlike other surfaces of magnetic elements, Fe, Co, and Ni,
where moments are strongly enhanced.

\begin{figure}
  \centering
  \includegraphics[width=\columnwidth]{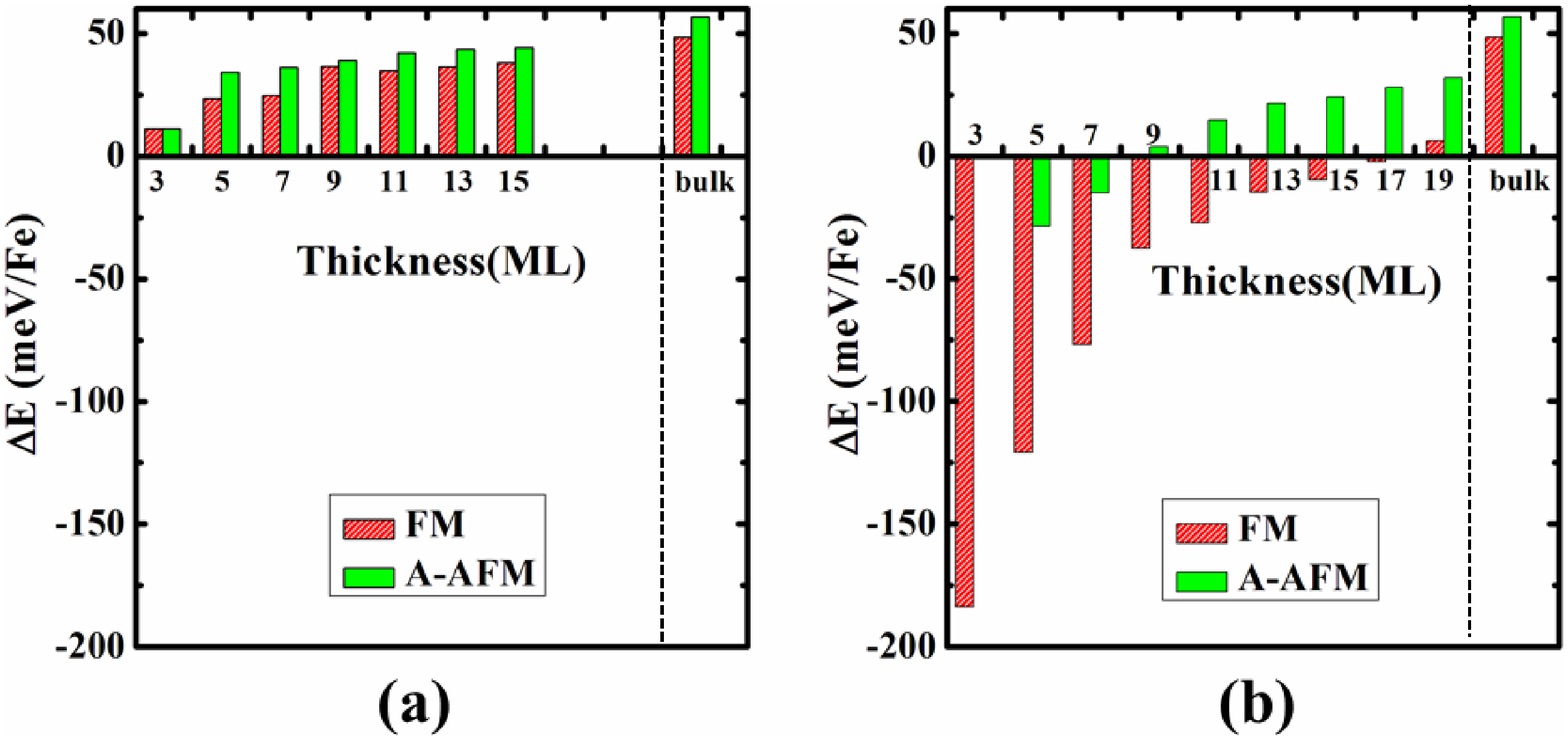}
  \caption{The thickness dependent total energy differences of the FM and A-AFM 
states from that of the G-AFM state, $\Delta E  = E_{FM} (E_{A}) - E_{G}$ ; (a) the Fe-terminated and (b) the Rh-terminated 
films.}
\label{fig:2}
\end{figure}
Before we discuss magnetism the FeRh films, we provide here a detailed analysis of bulk magnetism. 
$e_g$ and $t_{2g}$ are degenerate in G-AFM and FM due to the cubic symmetry,
whereas those in A- and C-AFM are no longer degenerate owing to the tetragonal magnetic structure (see SI Fig. S2).
Strong hybridization between Fe and Rh $d$ states brings in almost fully occupied (unoccupied) majority (minority) spin states
of Fe $d$ orbital, particularly $e_g$ states for all magnetic phases,
which results in enhanced moments of Fe atoms.
The nearly half-filled band favors AFM as in bulk Mn and Cr.
It is noteworthy that the majority spin bands of Rh are almost fully occupied in the FM similarly to Fe,
but featureless in AFM states.

The magnetic phase of FeRh alloy can be explained by interplay between three mechanisms 
- superexchange interaction\cite{20,21}, Zener-type direct-exchange interaction\cite{22}, 
and magnetic energy gain. In the framework of GKA rule, whether FM or AFM is preferred 
by superexchange interaction is explained by magnetic ion-ligand-magnetic ion angle. 
Here we view Fe atoms as magnetic ions and delocalized $s$ and $p$ 
orbitals of Rh as ligand orbitals in GKA rule. Fig. 1(d) schematically illustrates 
magnetic interactions between Fe atoms to be involved in determining magnetic structure. 
The GKA superexchange interactions are shown as solid lines and Zener-type direct interaction 
as a dotted line. In accord with GKA rule on magnetic coupling, Fe$_{1}$ 
prefers AFM coupling to Fe$_{2}$ and FM coupling to Fe$_{3}$ 
because angles of Fe$_{1}$-Rh-Fe$_{2}$ and Fe$_{1}$-Rh-Fe$_{3}$ 
are 180$^{\circ}$ and 109.5$^{\circ}$ (which is close to 90$^{\circ}$), 
respectively. 

On the other hand, the interaction between Fe$_{1}$ and Fe$_{4}$ 
is more or less direct since Rh atom is not much involved in this coupling. The 
half-filled $e_g$ states directing along the principal 
axes are more involved in the Zener direct interaction between Fe$_{1}$ 
and Fe$_{4}$ compared to the $t_{2g}$ states, 
which results in AFM coupling\cite{22}. Since the $d$ states are highly localized 
giving little wave function overlap, the Zener-type direct interaction between Fe$_{1}$ 
and Fe$_{4}$ must be weak. As a result of combination of the superexchange 
and the direct interactions discussed above, G-AFM is most stable among other AFM 
states. In the FM states, on the other hand, there is magnetic energy gain due 
to the considerable magnetic moment of Rh atom (1.041 $\mu_{B}$), 
which reduces total energy to a certain degree, hence a FM state is more stable 
than A-AFM and C-AFM states even though it is less stable compared to the ground 
G-AFM state. 

In the Rh-termination, the 180$^{\circ}$ superexchange interaction disappears 
because of the absence of Fe layer above the Rh-terminated surface.
The 90$^{\circ}$ superexchange interaction makes the A-AFM more stable than G-AFM.
Instead, the magnetic energy gain of the surface Rh atom plays a key role in stabilizing FM. 
Hence, FM is the magnetic ground state in the Rh-terminated FeRh thin films. In 
the Fe-termination, on the other hand, G-AFM is most stable as in bulk since the 
180$^{\circ}$ superexchange interaction still works. 
\begin{figure}
  \centering
  \includegraphics[width=\columnwidth]{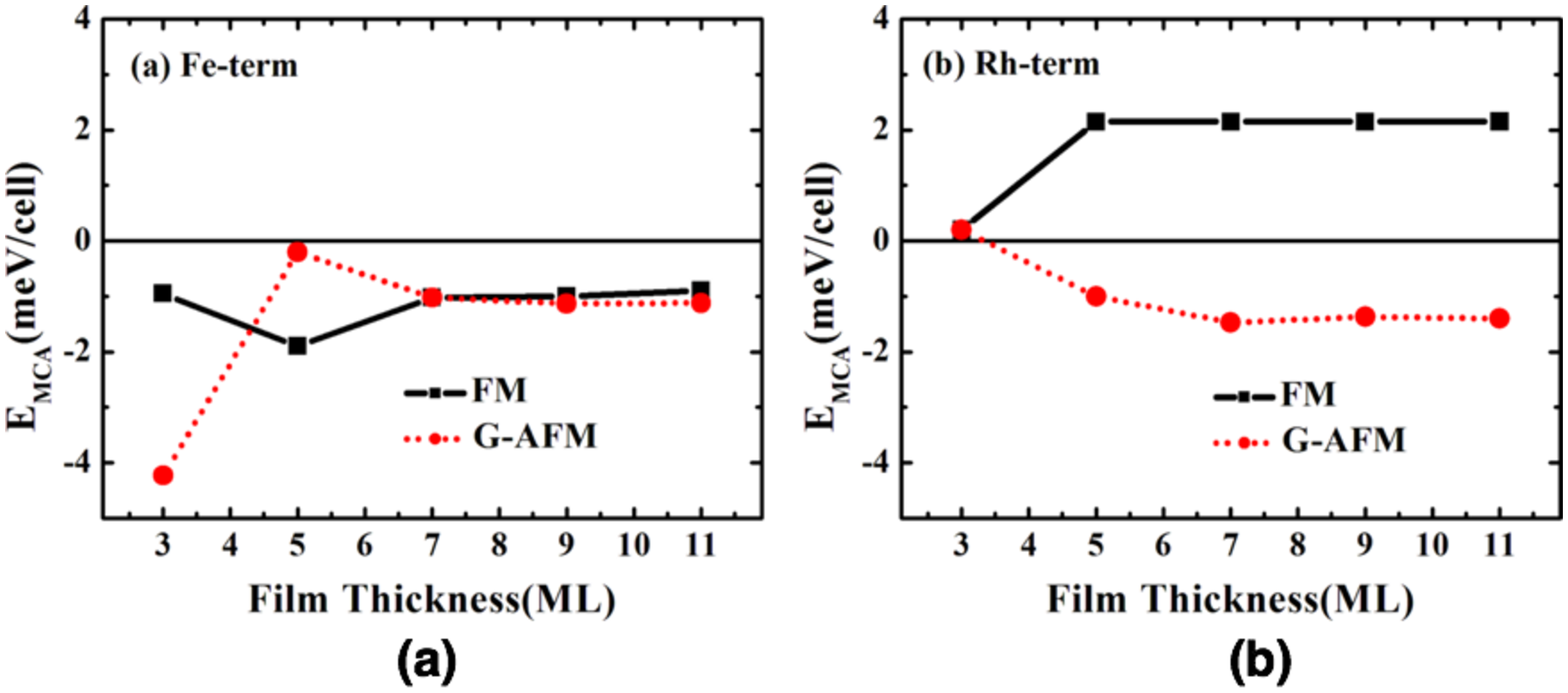}
  \caption{MCA energies, defined as $E_{MCA}\equiv E(\rightarrow)-E(\uparrow)$,
    of the FeRh (001) thin films for
    (a) Fe- and (b) Rh-terminated films, respectively.
    Solid (dashed) lines denote MCA of FM (G-AFM) phase.}
\label{fig:4}
\end{figure}

In Fig. 3, calculated MCA energies, $E_{MCA}\equiv E(\rightarrow)-E(\uparrow)$, are 
presented as function of thickness for the Fe- and the Rh-terminated films in their 
for G-AFM and FM states. 
$E_{MCA}$'s for other magnetic states are also shown for comparison. 
From the definition of $E_{MCA}$, positive (negative) value implies 
perpendicular (parallel) magnetization to the surface normal. Interestingly, the 
Rh-termination in FM state shows quite strong persistent perpendicular MCA regardless 
of thickness, whereas the Fe-termination exhibits in-plane MCA for all magnetic 
states. In particular, $E_{MCA}$ =+2.1 meV/$\Box$ of the Rh-termination 
is greater than 3$d$ magnetic metals such as bulk Fe, Co, and Ni by two orders of magnitude
and larger than magnetic films by several factors.
From the fact that both the Fe-terminated FM states and the Rh-terminated G-AFM state show
parallel MCA, the magnetic states are not a key factor in determining MCA, 
but the surface termination is.
Moreover, the saturated feature of $E_{MCA}$ 
=+2.1 meV/$\Box$ with respect to the thickness implies that magnetism of the Rh surface 
plays a key role in determining the strong perpendicular MCA. 
\begin{figure}
  \centering
  \includegraphics[width=\columnwidth]{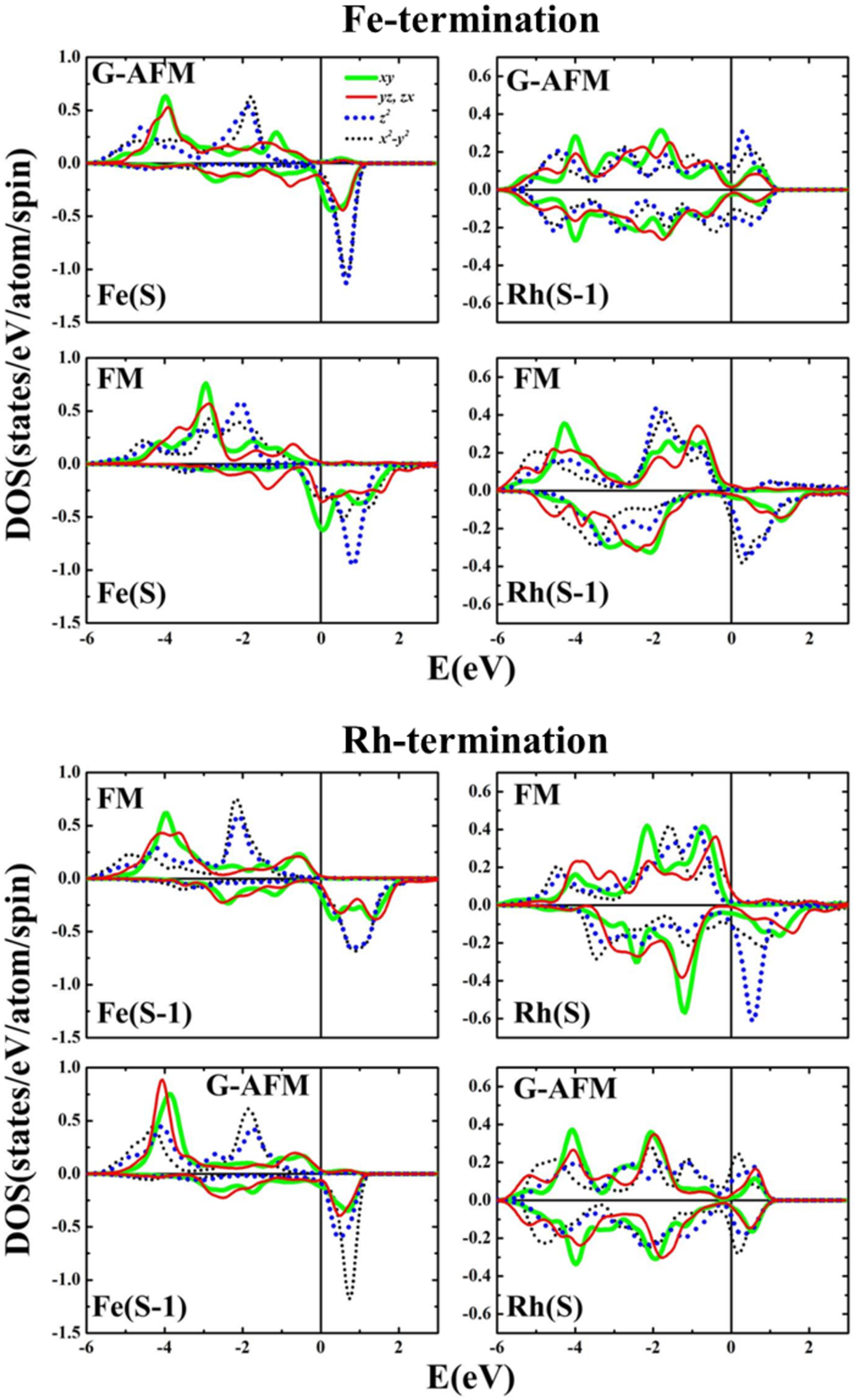}
  \caption{Projected density of states of $d$ orbitals of 9-ML FeRh film
    in the Fe-termination (upper panels) and the Rh-termination (lower panels) 
    for FM and G-AFM states.
    $S$ and $S$-$1$ indicate the top surface and subsurface layer, respectively.}
\label{fig:5}
\end{figure}
To discuss the role of the thickness and the surface termination on magnetism as well 
as MCA, we present density of states (DOS) of the Fe- and the Rh-terminated 9-ML film is presented 
in Fig. 4, where left and right columns represent DOS's of Fe and Rh atoms for better 
comparison as in Fig. S2.
First and second rows are for the Fe-terminated film in G-AFM and FM states;
third and forth rows are for the Rh-terminated film in FM and G-AFM states.
DOS features do not differ very much from the bulk,
despite lifted degeneracies in $e_{g}$ and $t_{2g}$ states at the surfaces
and some changes in surface atoms.
In particular, DOS's of subsurface atoms are essentially the same as the bulk.
This result confirms that the surface atoms indeed play the key role in determining magnetism in film geometry, 
as saturated $E_{MCA}$ with respect to thickness implies [see Fig.3].

To confirm and analyze calculated results on MCA, additional calculations have 
been carried out using FLAPW method using GGA for 5-, 7-, and 9-ML Rh-terminated films. 
Results by FLAPW method are quite consistent with VASP method: 
i) FM is much more stable than G-AFM states,
ii) magnetic moments of the surface Rh 
atoms are 1.107, 1.106, and 1.107 (Compare with SI Table S2) for 5-, 7-, and 9-ML, respectively,
and iii) $E_{MCA}$ =+1.97, +2.41, and 2.10 meV/$\Box$ (cf. Fig. 3), respectively. 
$E_{MCA}$'s are further decomposed into individual atomic contribution 
and different spin-channels\cite{31}. It is found that the surface Rh atom with stronger 
spin-orbit coupling\cite{32,33,34} and
the $\uparrow\downarrow$ channel play a dominant role in determining MCA.
Results using FLAPW adopting 
conventional von Barth-Hedin local spin density approximation (LSDA) are also listed 
in SI Table S3.

The role of the surface Rh layer in determining perpendicular MCA is well manifested in DOS:
the lifted degeneracy of $e_g$ states makes peak from $d_{z^2}$
more prominent in the minority spin band, which is just above the Fermi level. 
This peaked $d_{z^2}$ state contributes significantly to perpendicular MCA through
$\langle yz/zx,\uparrow|L_{X}|z^2,\downarrow\rangle$ matrix\cite{31} in the Rh-termination. 

In order to elucidate the magnetic behavior when the FeRh(001) films are under 
an applied magnetic field or heat, we present total energy of the Fe-terminated 
FeRh(001) thin film as a function of angle between magnetic orientations of the 
closest Fe atoms.[See SI Fig. S3]. The energy barrier of the AFM-to-FM transition 
under an applied magnetic field is reduced with decreasing of thickness. This information 
might be useful in designing spintronics devices such as an AFM memory\cite{8,9,10} and 
HAMR\cite{5,6,7}. 

In summary, magnetism of the Fe- and the Rh-terminated FeRh (001) are studied for 
various film thicknesses. The origin of stability of G-AFM in bulk is well explained 
in the framework of Goodenough-Kanamori-Anderson rules on the super-exchange interaction, 
where subsidiary Zener-type direct exchange interaction and magnetization energy are also 
taken into account. The thickness and the surface termination turn out really significant 
as the two terminations give different magnetic ground state. The Fe-termination 
is stabilized in G-AFM as in bulk, while the energy difference between G-AFM and 
FM is greatly reduced. On the other hand, the Rh-termination strongly prefers FM 
when films are thinner than 15-ML. Furthermore, the Rh-termination in FM state 
exhibits quite strong perpendicular MCA, 2.10 meV/$\Box$ ,
which is greater than bulk 3$d$ convention magnetic metals by two orders of magnitude 
and larger than 3$d$ magnetic films by several factors.
Utilizing the thickness and the surface-termination 
dependent magnetism of FeRh film, highly desirable materials can be tailored as 
required by spintronics devices. 

\begin{acknowledgments}
This work was supported by the Priority Research Centers Program (2009-0093818) 
and the Basic Science Research Program (2010-0008842) through the National Research 
Foundation funded by the Ministry of Education of Korea. ABS acknowledges support 
from the Czech Rep. grant (GACR 14-37427G).
\end{acknowledgments}


\end{document}